# IMPROVING RISK MANAGEMENT BY USING SMART CONTAINERS FOR REAL-TIME TRACEABILITY

*S. WATTANAKUL\*, S. HENRY\*\*, L. BENTAHA\*, N. REEVEERAKUL\*\*\*, Y. OUZROUT\**
*\* University of Lyon, University Lumière Lyon 2, DISP Laboratory, France*
*\*\* University of Lyon, University Lumière Lyon 1, DISP Laboratory, France*
*\*\*\* College of Arts, Media and Technology, Chiang Mai University, Thailand*
*Email: siraprapa.wattanakul@univ-lyon2.fr*

## Abstract

This research proposes implications of application functions by using the chain traceability data acquired from the *Smart Object attached with Extended Real-time Data* (*SO-ERD*: e.g. smart container, smart pallet, etc.) to improve risk management at the level of the logistics chain. Recent applications using traceability data and major issues in traceability systems have been explored by an academic literature. Information is classified by the usage of current traceability data for supporting risk detection and decisions in operational, tactical, and strategical levels. It is found that real-time data has been a significant impact on the usage for the transportation activity in all decision levels such the function of food quality control and collaborative planning among partners. However, there are some uncertainties in the aggregation of event-based traceability data captured by various partners which are preventing the adoption of data usage for the chain. Under the environment of Industry 4.0 and the Internet of Things (IoT), the SO-ERD enables independent data tracing through the chain in real-time. Its data has potential to overcome current issues and improve the supply chain risk management. Therefore, Implications of risk management are proposed with the usage of SO-ERD data based on the literature review which reveals current concerns of decision functions in the supply chain. The implications can be an impact to the domain needs.

**Keywords:** Risk Management, Chain Traceability, Internet of Things (IoT), Smart Container, Real-time Traceability

## 1. Introduction

Traceability is considered as an essential part of quality requirements to fulfil the quality improvement for organizations including in the supply chain. Traceability data collected during operations in the field is integrated for risk management in various contexts: manufacturing, food, pharmaceutical, etc. [2,3]. However, current traceability practice is limited by information visibility within the chain which consequently impacts the supply chain risk management (SCRM).

To track and trace products among stakeholders along the supply chain, Electronic Product Code Information Services (EPCIS), is the only global standard which uses barcode and/or RFID technologies to perform four major activities for the traceability: (1) to identify traceable items, (2) to capture events (date-time, location, event-type e.g. arrival, transfer, etc.) separately by each partner using their own barcode or RFID reader, (3) to share business-critical information with involved partners and (4) to trace the product flow among trading partners [4]. However, the frequency of event capturing limits the visibility of event-based traceability. The tracing status in between reading points during logistics is invisible (as shown in Figure 1) which further restricts the risk awareness and risk handling. Moreover, due to the scattering of data storage owned by different stakeholders, it requires extra efforts to qualify chain data. The additional technical implementation at least the data crawling to fulfil requirements of the chain collaboration [5].

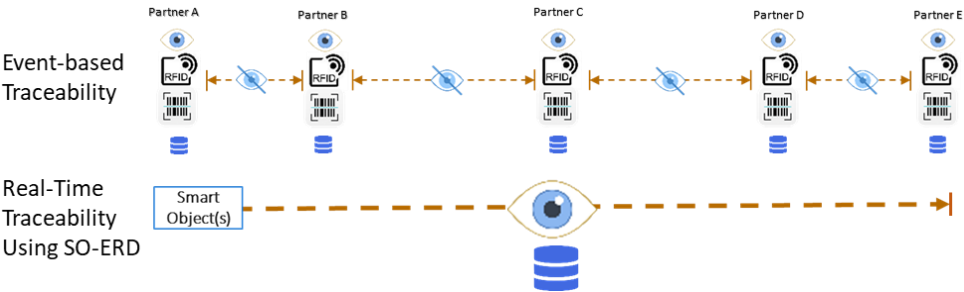

Figure 1: The visibility of EPCIS Traceability and Real-Time Traceability through the supply chain.

Soon within the Industry 4.0 environment, the visibility is extending through the chain during transportation in real-time by technologies of the Internet of Things (IoT). For example, Traxens proposes a smart container solution which can sense the actual status of the tracing object (e.g. temperature, container's status, location, etc.) and collect these data to one global data storage in near real-time, not limited to the event-based. Traxens also offers data service platform for clients to monitor smart container and related data in real-time [6]. These data collected by the *Smart Object attached with Extended Real-time Data* (SO-ERD) has potential to extend the use for SCRM but current studies are limited.

Therefore, this paper proposes the use of SO-ERD data to improve SCRM as compared to the conventional methods in the literature review. Besides, the SO-ERD traceability data can be used to handle risks in the supply chain level and overcome difficulties of the current traceability system. The rest of the paper is organized as follows. Section 2 presents the theoretical background of the risk management and definitions of traceability. Section 3 describes the methodology of the academic literature review. Section 4 reveals results of the usage of traceability data to support decisions on the supply chain. Section 5 introduces implications of the chain traceability data captured by SO-ERD. Finally, Section 6 concludes the paper.

## 2. Theoretical background
### 2.1 Risk Management

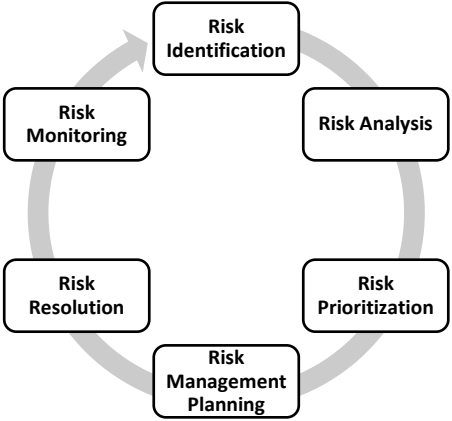

Figure 2 introduces a typical Risk Management cycle [7], (1) identifying risk items which tends to be success's obstacles, (2) assessing loss probability and loss magnitude of each risks e.g. techniques of performance models, (3) prioritizing the order of risk items, (4) planning of risk management to prepare risk handling, (5) performing resolution to the risk and finally (6) monitoring to progress after resolving and continuously managing risks. To perform risk management activities, it is necessary to have the available data for the monitoring process. Especially, risk identification and analysis depend on data captured during operations in the field and are captured by traceability tools.

Figure 2: Risk Management Cycle [7]

### 2.2 Definitions of Traceability

ISO 9000:2015 defines traceability as "ability to trace the history, application, or location of an object" while "object" means "anything perceivable or conceivable" such as product, process, service, person, system, organization, or resource. Traceability of a product or a service can also include the processing record, the distribution and location after delivery [1]. Traceability is also classified in terms of partnership by Moe [8] as follows:

- *Internal Traceability*: tracing within an organizational internally, e.g. batch processing.
- *Chain Traceability*: tracing the continuous production along with several partners which go from materials or resources, productions, distributors and/or others. It can be conducted in two different approaches for chain data collection as illustrated in Figure 3.
  - Approach 1: Distributed - Only the identity of the product is passed through the chain, information can be acquired by collaborating with internal traceability.
  - Approach 2: Accumulated - All information is attached to the product along with all stages of the chain.

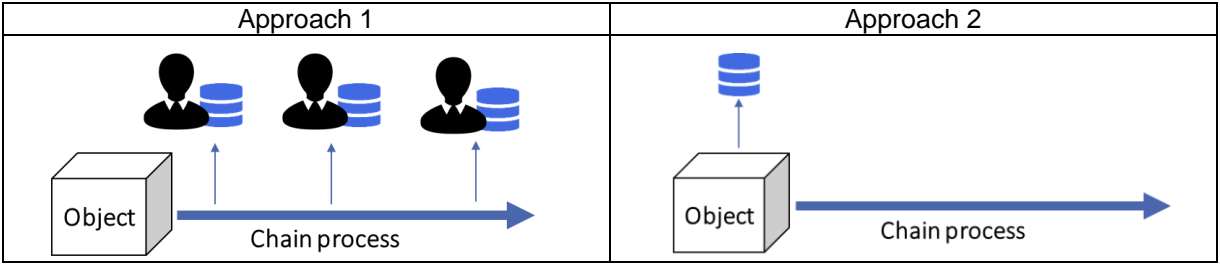

Figure 3: Chain Traceability approaches

Currently, the common approach for chain traceability is Approach 1 as such the practice of EPCIS standard. For this approach, as an example, in the case of a container traceability, this later travels through several logistics service providers. When passing on each provider, internal operations can collect information via RFID or barcode. Tracing of the container through the whole chain requires collaborations among these partners to aggregate all data. While Approach 2 is scarce to be adopt using technology due to it is difficult to attach data to an object during the travel. However, the availability of IoT which facilitates objects to extract information data anywhere can overcome this difficulty. The adopting of Approach 2 becomes possible for the case of smart container. It can collect data by itself when it travels through service providers of the chain. The collaboration to collect the internal traceability of logistics data from partners is not mandatory. Therefore, the approach of chain traceability for the SO-ERD turns to the Approach 2, as well as tasks affected by traceability data such as the risk management.

## 3. Methodology

In order to identify opportunities for the adoption of risk management, decision making is a significant activity to identify and to analyse risk. Therefore, this study focuses to make use of the traceability data to support decisions of the supply chain. A literature review is conducted within main bibliography sources; Scopus, Web of Science and IEEE. First, major related keywords are searched, "traceability" and "data", and "supply chain" or "logistics", and "decision" or "performance". After merging results from all databases to remove duplicated papers, there are 142 publications from the year 1996 to 2017. Then, papers are reviewed to exclude studies which have no usage of traceability data or the usage is only for tracking and tracing or high-level usage concept without data parameters, for example, studies of methods for tracing resource originality, studies of stakeholder's perspective based on survey, studies that adopt technologies such as RFID for traceability, conceptual papers and etc. The remained 31 publications [10-40] are reviewed again in detail and information is structured using mind mapping.
The classifications on mind map are structured in three major axes.

1. *Traceability Data* – containing a list of data parameter and type of the data (e.g. master data, transport condition, business transaction), the frequency of traceability data acquisition: real-time/ event based/ batch/ offline and the acquisition tools.
2. *Usage of Traceability Data* f*or Decision Support* – the use of traceability data is classified into three levels: operational, tactical and strategical. The impact of the decision is also classified into activities of the supply chain process with supported data attributes.
3. *Issues* – presenting issues of technical and business perspectives caused by implementing current traceability system.

## 4. Result

Regarding 31 publications researching in the use of traceability data are in the context of perishable products, manufacturing, transportation service providers and others in the ratio of 54.8%, 25.8%, 9.7% and 9.7%, respectively. Details of studies are classified into three axes as follows:

## 4.1 Traceability data

Traceability data is classified into three types by its characteristics: Master, Transactional and Condition Status. Master data is containing permanent or not subject to frequent change data such as identifications, product information, location information, etc. Transactional data is containing events captured during the flow of tracing object e.g. shipment information (date of dispatch/arrival, to/from, delivery operator). Condition Status captures parameters related to the tracing object. This last data type can be the status of surrounding environment, e.g. temperature, humidity, position, etc or the status of tracing object's component, e.g. nutrients of food product occurring during the production in the field. In terms of data acquisition, the visibility of the physical activities occurring in the process flow depends on the availability of data and information. A data acquisition is conducted by different tools e.g. RFID, sensor, GPS, Laboratory or Manual. Data is captured in different frequency: real-time, batch or event-based. For the real-time capturing, data is detected and transmitted immediately and continuously to the information system. For the batch capturing, data is collected once in a time interval or once in a production lot. For an event-based, data is collected by an event action such as product arrival/depart time, recall, etc. The classification of data acquisition of traceability data is as presented in Table 1.
In a traceability system, it uses a combination set of data types. Master data is statically defined in the information system and is available before productions in the field is launched. From the review studies, 67.7% of applications uses the Transactional data and 71.0% of applications uses the Condition Status data. The number of studies in Conditional Status data is a little higher even there is no traceability standard supporting. Especially the use of data attributes in temperature, position and humidity are in 38.7%, 29.0%, 16.1% of total research respectively.

| Data Acquisition | Transactional | Conditional Status | | | | | |
|---|---|---|---|---|---|---|---|
| | | Temperature | Humidity | Position | Gas | Object's component | Others |
| *Classification by Acquisition tool* | | | | | | | |
| RFID/barcode | [11,12,15,18,19,21,22,25,27,28,29,30,31,36,37,39] | | | [15,19,21,27,28] | | | |
| Sensor | | [10,11,12,13,16,23,24,25,26,29,32] | [10,12,24,26,29] | | [10] | | [13] |
| GPS | | | | [18,22,30] | | | |
| *Manual/Laboratory* | [17,34,40] | [20] | | | | [20,13,35] | [20,40] |
| *Classification by Acquisition frequency* | | | | | | | |
| Real-time | | [10,11,12,16,23,24,26] | [10,12,24,26] | [14,22,30] | [10] | | |
| Batch of production or time interval | [25] | [13,20,25,32] | | | | [13,20,35] | [13,20] |
| Event-based | [11,12,14,15,17,18,19,21,22,27,28,29,30,31,33,34,37,38,39,40] | [29] | [29] | [15,18,19,21,27,28] | | | [40] |

Table 1: The classification of data acquisition of traceability data in tools and frequencies

## 4.2 Usage of traceability data

Once the traceability data is acquired, it is prompt to support activities of the supply chain process. This study classifies the use of traceability data supporting on each decision level: in operational – decisions during the flow of tracing object, in tactical – decisions before the operation activity e.g. planning, scheduling etc., and in strategical – decisions in a process or business change. Furthermore, to identify the impact of traceability data, the usage of decisions is mapped to the 9 activities of supply chain process defined by Grant et al. [9]. The classification result of traceability data usage is as shown in Table 2.

| Usage of data | Decision Support Level | | |
|---|---|---|---|
| | Operational | Tactical | Strategical |
| *Classification by frequency of data analysis* | | | |
| Real-time | [10,11,12,16,22,23,24,26,30,32] | [14,22,30] | |
| Event-based | [12,19,25,27,28,29,31,40] | [15,18,28] | [12,37] |
| Batch of production | [13,20,35] | [31] | |
| End of cycle | | [17] | [33,36,38] |
| Series of process cycle | | [21] | [34,39] |
| *Classification by Impacted activities* | | | |
| 1. Customer service and support | [11] | [14] | [33,34,36,37,38,39] |
| 2. Demand forecasting and planning | | [14,21] | [33,34,36,37,38,39] |
| 3. Purchasing and procurement | | [14] | [34] |
| 4. Inventory management | | [14,31] | [34] |
| 5. Order processing and logistics communications | [10,12,13,20,23,28,31,35,40] | [15,18,21,28] | [12,34] |
| 6. Material handling and packaging | [10,12] | | [12,39] |
| 7. Transportation | [10,11,12,13,16,22,23,24,25,26,27,29,30,32] | [14,17,18,22,30] | [12,39] |
| 8. Facilities site selection, warehousing and storage | [10,12,13,16,19,20,23,24,29,32] | | [12,39] |
| 9. Return goods handling and reverse logistics | [29,35] | | |

Table 2: Usages of traceability data for decision support and performance measurement in the supply chain

The total 31 studies in decisions: 20 studies in operational, 9 studies in tactical and 7 studies in strategical. The most usage of data is in the transportation activity for 54.8% of the total. Results are as follows:

### 4.2.1 Decisions in Operational

The highest number of traceability data usages are in the activities of processing, transportation, and storage. Functions of this decision level are generally concerning in the awareness of risks before they occur such as anomaly prediction, pre-warning prediction food shelf-life prediction and fault diagnosis. Therefore, real-time analysis is significant. From the top mentioned activities, about half of the applications use real-time data for decisions. Furthermore, every application of real-time decisions in operational and tactical (Table 2) need to use the traceability data captured in real-time (Table 1) to support their real-time analysis.

### 4.2.2 Decisions in Tactical

In this decision level, the '*position*' parameter plays an important role to support planning and scheduling of the manufacturing [15,21,28] and to improve collaboration among partners [14,18,22,30]. The '*position*' can be used for acknowledging the position of shared object or can be mining for the knowledge of object trajectory to support plan optimization.

### 4.2.3 Decisions in Strategical

Traceability data is used as secondary information to support the strategical decision. Production data is modelled to illustrate the big picture of the process. The first usage is to measure the performance of the operation process. A method is by integrating the Business Process Management Notation (BPMN) to the process interaction model. KPIs are measured and analysed for each activity performance along the chain process [12]. In the case of environmental performance, traceability graph is mapped to environmental KPIs e.g. $CO_2$, $CH_4$, $N_2O$ etc. [33,36,38]. The second usage is to model process from historical data for simulating the process re-engineering. The process model is constructed for the experiment of process change before the actual implementation [34,39].

## 4.3 Issues of chain traceability implementation

Regarding the implementation of traceability data usage, there are challenges to the current solutions which may prevent the adoption of the risk management system are described:

### 1. Traceability data sharing

Regarding that global traceability of event-based approach, only the relevant data captured by each partner should be shared while respecting confidentialities of all partners [18]. Only necessary interfaces among partners are limited. However, partners are still lack of willingness to share due to concerning in data security and reliability [12,14] and some partner can choose not to share access code of EPC on their sites [18].

Moreover, in addition to the selected 31 papers studying the use of traceability data, it is interesting that there are 14 papers which propose data models to aggregate data of the chain. These models tend to improve visibility and consistency for real-time tracing, collaboration, and decision making. By the number of these studies, it is about half compared to selected studies. It is assumed that currently there are requirements to improve the linkage of data and information for sharing in the chain level.

### 2. Changes in business

There can be changes in business which cause uncertainty to practice, process model [14] and decision-making procedure [22]. A dynamic information system which is flexible to changes and enables real-time feedback for precise adaptive decision is required [22].

## 5. Implications

Based on the capability of the SO-ERD, the theoretical background, and the literature review result, this study proposes implications toward the implementation of traceability data of SO-ERD for the risk management of supply chain.

### 1. SO-ERD offers the change of chain traceability approach.

Current chain traceability is performed by collecting event-based distributed data among the chain partners. SO-ERD offers opportunity in collecting data by itself through the logistics chain. Therefore, the chain traceability approach can be changed to the accumulated collection.

### 2. SO-ERD can overcome the data aggregation issue of distributed data collection.

As a result of the accumulated approach of the chain traceability, concerning in aggregating scattered data from the chain partners can be declined. The chain traceability data is also available in real-time.

### 3. SO-ERD is matched to the current research focus in the traceability data usages for the supply chain.

Once the availability of chain data is enable, the extension of the data value is possible. Through the interest of reviewed researches, activity during the transportation has the highest usage of traceability data especially the decision functions which use real-time data. The SO-ERD such as smart container also has capability to collect real-time data during the transportation. Therefore, the potential of using SO-ERD to serve the domain needs can be an impact. The implementation of usage functions based on SO-ERD data should be also matched to current requirements as stated by the literature review.

4. *SO-ERD offers real-time traceability data to support real-time analysis of all decision levels under the environment of Industry 4.0.*

As the SO-ERD independently collects data of itself at anytime and anywhere it goes in real-time but data is not limited just only to itself status. Under the environment of Industry 4.0, the integration of traceability data with external data sources such as weather, route condition, business information system, etc can extend information and enable the awareness of the situation. This information can support the decision analysis for transportation activity in real-time such the case of a smart container as follows:

- *Operational decisions: Robust Tracking* – smart container offers various data parameters such as position, temperature, humidity, etc. so not only the common function as the goods position tracking but the goods quality tracking can also be functioned. By these parameters, the analysis for problem detection such as the probability of container lost, accident or goods quality prediction can be conducted, notified and handled in real-time.

- *Tactical decision: Efficient Planning* – due to the availability of chain data in real-time, knowledge of the travelling route can be constructed in several dimensions by integrating the container data such as '*position*' and '*time*' with the external related data such as weather, traffic congestion, operator performance etc. Then, the prediction or optimization decision functions of the route plan or route collaboration can be more informative. In addition, as the data of the chain available to all parties, the internal analysis can also benefit the chain data such as the prediction of incoming loads for the port preparation based on the real-time situation.

- *Strategical decision: Process Performance* – the handling of containers by each operator are different. The measuring of time, position, vibration in container, etc can be used to evaluate the performance of each operator and can be conducted anytime. This information can support the chain for further process improvement analysis.

5. *The recursive monitoring of risk management for the logistics chain is possible by SO-ERD.*

Since the SO-ERD data is available in real-time, the risk management of the chain according to the risk management cycle can be performed at all decision levels recursively in real-time. During the transportation, risks are monitored, notified and analysed continuously for real-time handling. For tactical and strategical, the route plan can be decided based on the analysis of risk prevention or to achieve the optimized solution. Once the action is taken as planned, results of the plan or a process change can be feedbacked immediately throughout the chain.

6. *The recursive monitoring of risk management in real-time can improve robustness to the business or process change.*

Data acquired by SO-ERD is robust to changes in a process due to the data collection is independent of the process. Even there are changes in intentionally or unintentionally to business strategy, process or production, traceability data is monitored in real-time and can be notified in an early time.

## 6. Conclusions and Future Research

The major challenges for the chain traceability are the scattering of data collection and aggregation, and the uncertainty of business process. These have influenced on the SCRM due to the limitation of data availability. This paper proposes implications of the SO-ERD traceability data to extend the risk management of the logistics chain. The proposed decision functions are also based on the literature review presenting current requirements dealing with concerns of the supply chain. This might improve the chance of impact to the domain once these functions of SO-ERD data usage are implemented. In near future, we will propose design of the proposed decision functions for further improvement of the SO-ERD traceability.

## Acknowledgements
This work is supported under the Erasmus Mundus's SMARTLINK (South-east-west Mobility for Advanced Research, Learning, Innovation, Network and Knowledge) Project.